\documentstyle[aps,prl,multicol,epsfig]{revtex}
\newcommand{\bq}{\begin{equation}}
\newcommand{\eq}{\end{equation}}
\newcommand{\bqa}{\begin{eqnarray}}
\newcommand{\eqa}{\end{eqnarray}}
\newcommand{\nn}{\nonumber \\}

\begin{document}
\draft 
\title{Doping dependence of bose condensation energy and correlations with spectral peak intensity and superfluid weight in high $T_c$ cuprates}
\author{Sung-Sik Lee and Sung-Ho Suck Salk$^1$}
\address{Department of Physics, Pohang University of Science and Technology,\\
Pohang, Kyoungbuk, Korea 790-784\\
$^1$ Korea Institute of Advanced Studies, Seoul 130-012, Korea\\}
\date{\today}

\maketitle

\begin{abstract}
Based on our recent holon-pair boson theory of the t-J Hamiltonian (Phys. Rev. B {\bf 64}, 052501 (2001)) we report the doping dependence of the bose condensation energy, superfluid weight and spectral peak intensity.
We find a universality of doping dependence in these physical quantities, by equally showing an arch shape in the variations of their magnitudes with the hole doping concentration.
We find that all of these physical quantities scale well with the positive charge carrier (hole) density $x$, but not with the electron density $1-x$ for the entire range of hole doping.
It is shown that the doping dependence of the condensation energy $U$ at $T = 0 K$ is given by the relation, $U(0)  \approx \alpha x^2 |\Delta_0|^2$ with  $\Delta_0$, the pairing gap at $0 K$ and $\alpha$, a constant. 
\end{abstract}
\begin{multicols}{2}

\newpage
In conventional superconductors, condensation energy $U$ is well explained by the BCS theory based on the Fermi liquid picture, that is, $U = \frac{1}{2} N(0) \Delta^2_0$, where $N(0)$ is the density of single-electron state at the Fermi surface in the normal state and $\Delta_0$, the superconducting gap at zero temperature\cite{SCHRIEFFER}.
On the other hand, with increasing electron density $1-x$ or with decreasing hole density $x$ underdoped high $T_c$ cuprates have revealed an unusual doping dependence by showing a decreasing trend of the condensation energy $U$ and an increasing trend of zero temperature gap $\Delta_0$\cite{LORAM94,LORAM00,LORAM01,MOMONO}.
Most recently experimental studies of condensation energy are reported based on the specific heat measurements of $La_{2-x}Sr_{x}CuO_4$ by Momono et al.\cite{MOMONO} and Loram and Tallon\cite{LORAM94}.
Their estimations of the condensation energy from the specific heat measurements showed a good agreement in the overdoped region but not in the underdoped region.
Momono et al.\cite{MOMONO} reported from their electronic specific heat measurements that the observed condensation energy in the underdoped region does not correlate well with the conventional BCS condensation energy for the Fermi liquid and, thus, can not be explained only by the reduction of $N(0)$. 
Instead they were able to fit the observed condensation energy by introducing an effective superconducting gap $\Delta_{eff} = b x \Delta_0$ in place of $\Delta_0$ in the BCS formula\cite{MOMONO}.
Here $x$ is the hole doping concentration and $b$, a constant.
This implies a strong hole (positive charge carrier) dependence of the condensation energy.

There exist various levels of theoretical studies of condensation energy for high $T_c$ cuprates.
Earlier Scalapino et al.\cite{SCALAPINO} suggested that the condensation energy should be proportional to the change in the exchange energy between the superconducting state and the normal state if the antiferromgnetic interaction is responsible for the superconductivity.
Norman et al.\cite{NORMAN} pointed out that condensation energy is related to the changes in the spectral function between the superconducting state and the normal state. 
Most recently, based on the slave-boson approach to the $t-t^{'}-J$ model, Li et al.\cite{LI} showed that the condensation energy decreases with the increase of hole doping in the overdoped region.
Abanov et al.\cite{ABANOV} showed from the use of the spin-fermion model that the condensation energy is contributed from both the resonance peak at ${\bf Q}=(\pi,\pi)$ and a wide range of frequencies up to $J$.
Recently Feng et al.\cite{FENG} showed that both the superfluid weight and the superconducting peak ratio (SPR) have a similar doping dependence as the condensation energy.
Further, to the best of our knowledge there has been no theoretical report of comprehensive investigations paying attention to possible correlations between the condensation energy, superfluid weight and spectral function.
Presently, there exists a lack of comprehensive theoretical study dealing with a full range of doping dependence of the condensation energy which covers both underdoped and overdoped regions.

Earlier, based on our improved holon-pair boson theory\cite{LEE} we obtained the arch-shaped superconducting transition temperature as a function of hole concentration in agreement with the observed phase diagrams.
Further this theory predicted the peak (Drude peak)-dip-hump (mid-infraband) structure of optical conductivity below $T_c$ in agreement with observations\cite{LEE_OPT}.
In this work, employing the same theory, we present a comprehensive study of the condensation energy, the superfluid weight and the spectral peak intensity at $(\pi,0)$ and discuss our findings on correlations between these physical quantities and the universality of doping dependence of the bose condensation energy, the superfluid weight and spectral intensity.

We write the t-J Hamiltonian, 
\begin{eqnarray}
H & = & -t\sum_{<i,j>}(c_{i\sigma}^{\dagger}c_{j\sigma} + c.c.) + J\sum_{<i,j>}({\bf S}_{i} \cdot {\bf S}_{j} - \frac{1}{4}n_{i}n_{j}).
\label{eq:tjmodel1}
\end{eqnarray}
Here ${\bf S}_{i}$ is the electron spin operator at site $i$, ${\bf S}_{i}=\frac{1}{2}c_{i\alpha}^{\dagger} \bbox{\sigma}_{\alpha \beta}c_{i\beta}$ with $\bbox{\sigma}_{\alpha \beta}$, the Pauli spin matrix element and $n_i$, the electron number operator at site $i$, $n_i=c_{i\sigma}^{\dagger}c_{i\sigma}$.
Using the single occupancy constraint and thus $c_{i \sigma}  =  b_i^\dagger f_{i \sigma}$\cite{KOTLIAR,UBBENS} (with $f_{i \sigma}$, the spinon (spin $1/2$ with charge $0$) annihilation operator of electron spin $\sigma$ and $b_i^\dagger$, the holon (charge $+e$ with spin $0$) creation operator at site $i$), the U(1) slave-boson representation of the above t-J Hamiltonian is obtained to be
\bqa
H & = & -t \sum_{<i,j>}(f_{i\sigma}^{\dagger}f_{j\sigma}b_{j}^{\dagger}b_{i} + c. c.) \nn
&& -\frac{J}{2} \sum_{<i,j>}  b_i b_j b_j^\dagger b_i^\dagger (f_{\downarrow i}^{\dagger}f_{\uparrow j}^{\dagger}-f_{\uparrow i}^ {\dagger}f_{\downarrow j}^{\dagger})(f_{\uparrow j}f_{\downarrow i}-f_{\downarrow j}f_{\uparrow i}) \nn
&& + i\sum_{i} \lambda_{i}(f_{i\sigma}^{\dagger}f_{i\sigma}+b_{i}^{\dagger}b_{i} -1),
\label{eq:u1_sb_representation}
\eqa
where $\lambda_{i}$ is the Lagrange multiplier field to enforce the single occupancy constraint at each site.

The second term of the above equation shows the coupling between spin(spinon) and charge(holon) degrees of freedom.
Let us take a look at the necessity of retaining the charge degree of freedom.
The Coulomb repulsion term in the Hubbard Hamiltonian is decomposed into $U n_{i\uparrow} n_{i\downarrow} = \frac{U}{4}(n_{i\uparrow} + n_{i\downarrow})^2 - \frac{U}{4}(n_{i\uparrow}-n_{i\downarrow})^2$, representing both the charge (the first term) and spin (the second term) degrees of freedom. 
It should be noted that the Hubbard Hamiltonian is mapped into the t-J Hamiltonian in the large $U$ (Coulomb repulsion) limit. 
Thus the neglect of the charge degree of freedom in the interaction term of the t-J Hamiltonian is physically incorrect.
The Heisenberg term ${\bf S}_{i} \cdot {\bf S}_{j} - \frac{1}{4}n_{i}n_{j}$ leads to  the coupling between the charge (holon pairing) and spin (spinon pairing) degrees of freedom as shown by $-\frac{1}{2} b_i b_j b_j^\dagger b_i^\dagger (f_{i \downarrow}^\dagger f_{j \uparrow}^\dagger - f_{i \uparrow}^\dagger f_{j \downarrow}^\dagger ) ( f_{j \uparrow} f_{i \downarrow} - f_{j \downarrow} f_{i \uparrow} )$.
In the literature such charge contribution is often neglected.
The uncertainty principle exists between the number density (amplitude ) and the phase of a boson order parameter. 
For charged bose particles like the Cooper pairs, the number density fluctuations imply the fluctuations in charge density. 
For short coherence length superconductors, the local (or short range) charge density fluctuations need to be taken into account.
Thus, not only the phase fluctuations but also the local charge fluctuations should be admitted for high $T_c$ superconductors as manifestations of quantum fluctuations .
In the present study we retained such charge (holon) part of contributions.

After Hubbard-Stratonovich transformations and saddle-point approximation, we obtain the mean field Hamiltonian for the U(1) theory\cite{LEE}, $H^{MF}= H_{\Delta,\chi} + H_f + H_b$, where we have for the saddle point contribution of the order parameters,
\bqa
H_{\Delta,\chi} & =& \sum_{<i,j>}\Bigl[ \frac{J^{'}}{2} |\Delta_{ji}^{f}|^{2} + \frac{J^{'}}{4} |\chi_{ji}|^{2}  \nn
&& + \frac{J}{2}|\Delta^f_{ij}|^2|\Delta_{ji}^{b}|^{2} +\frac{J}{2} |\Delta^f_{ji}|^2 x^2 \Bigr],
\eqa
for the spinon sector,
\bqa
H_f & =& -\frac{J^{'}}{4} \sum_{<i,j>,\sigma} \Bigl[ \chi_{ji}^{*} (f_{j\sigma}^{\dagger}f_{i\sigma}) + c.c. \Bigr] \nn
&& - \sum_{i,\sigma} \mu^{f}_{i} \left( f_{i\sigma}^{\dagger} f_{i\sigma} -(1-x) \right) \nn
&& -\frac{J^{'}}{2} \sum_{<i,j>} \Bigl[ \Delta_{ji}^{f*} (f_{j\uparrow}f_{i\downarrow}-f_{j\downarrow}f_{i\uparrow}) + c.c. \Bigr],
\label{eq:spinon_u1}
\eqa
and for the holon sector,
\bqa
H_b & = & -t \sum_{<i,j>} \Bigl[ \chi_{ji}^{*}(b_{j}^{\dagger}b_{i}) + c.c.  \Bigr] -\sum_{i} \mu_{i}^{b} ( b_{i}^{\dagger}b_{i} -x ) \nn
 && -\sum_{<i,j> } \frac{J}{2}|\Delta^f_{ij}|^2 \Bigl[ \Delta_{ji}^{b*} (b_{i}b_{j}) + c.c. \Bigr],
\label{eq:holon_u1}
\end{eqnarray}
where $J^{'}=(1-x)^2J$, $\chi_{ji}= < f_{j\sigma}^{\dagger}f_{i\sigma} + \frac{4t}{J^{'}} b_{j}^{\dagger}b_{i}>$, 
$\Delta_{ji}^{f}=< f_{j\uparrow}f_{i\downarrow}-f_{j\downarrow}f_{i\uparrow} >$, 
$\Delta_{ji}^{b} = <b_{j}b_{i }>$, and $\mu^{f}_i$($\mu^{b}_i$), the spinon (holon) chemical potential.
We obtain the free energy for the U(1) theory\cite{LEE},
\begin{eqnarray}
\lefteqn{F  =   NJ^{'} \Bigl( \Delta_f^{2} + \frac{1}{2}\chi^{2} \Bigr) }\nn
&& - 2k_{B}T \sum_{k} ln [ \cosh (\beta E_{k}^{f}/2) ] - Nx \mu^{f} - 2Nk_{B}Tln2  \nn
&& + NJ\Delta_f^2 (\Delta_b^{2}+x^2) + k_{B}T \sum_{k} ln [1 - e^{-\beta E_{k}^{b}}] \nn
&& + \sum_{k} \frac{ E_{k}^{b}+\mu^{b}}{2} + N x \mu^{b},
\label{eq:u1_free_energy}
\end{eqnarray}
where $E_{k}^{f}$ and $E_{k}^{b}$ are the quasiparticle energies of spinon and holon, respectively.
Here $\beta = \frac{1}{k_B T}$ and $N$, the number of sites.

The condensation energy is the energy difference between the normal state and the superconducting state at $0K$.
In the present theory, the superconducting state is characterized by the presence of hole-pair bose condensation as a result of coupling between the holon pair and the spinon pair.
Thus the condensation energy density is given by
\bqa
U = \frac{1}{N} \left[ F_N( \chi, \Delta_f, \Delta_b = 0 ) - F_S( \chi, \Delta_f, \Delta_b ) \right].
\eqa
Here, $F_N$ is the free energy of the normal state obtained from setting $\Delta_b = 0$ and $F_S$, the free energy of the superconducting state.
With the choice of $J/t=0.2$ Fig. 1 displays the doping dependence of condensation energy density (solid circle) at zero temperature and the doping dependence of superconducting transition temperature $T_c$ (open circle) respectively.
Both cases showed an arch shaped variation of their magnitudes as a function of hole doping as shown in the figure.
The condensation energy is shown to increase with increasing hole concentration up to a critical value $x_c$ which occurred above the optimal doping $x_o$ (predicted at $0.07$).
Such trends are in complete agreement with observations.
However, we can state that only qualitative agreements with observations\cite{LORAM94,MOMONO} are achieved since there exists a quantitative shortcoming of predicted optimal doping value.

Here we discuss the physics involved with the condensation energy.
The last term in Eq.(\ref{eq:holon_u1}) represents the coupling of the holon pairing ($\Delta_b$) channel to the spinon pairing ($\Delta_b$) channel which causes to form the Cooper pairs.
Thus this simple coupling term holds a key.
It is readily understood from this coupling term that the Cooper pair order parameter diminishes when the spinon pairing order parameter becomes smaller in the heavily overdoped region. 
Indeed, in agreement with observations\cite{LORAM94,MOMONO}, the computed condensation energy showed an arch shape as a function of hole doping concentration by exhibiting a diminishing trend in the overdoped region.
In order to see correlations with the condensation energy, we computed the the mean field energy of $\frac{J}{2} |\Delta_f|^2 |\Delta_b|^2$ corresponding to the last term in Eq.(\ref{eq:holon_u1}).
The magnitude of the condensation energy scales well with the mean field energy of the hole pair (a composite of the spinon pair and the holon pair) term (the last term in Eq.(\ref{eq:holon_u1})) in that they have nearly the same order of magnitude variation over the most range of hole doping.
The magnitude of condensation energy increases with the Heisenberg exchange coupling strength $J$ at all doping concentrations $x$.
Based on the combinatory study of both the $x$ and $J$ dependence of condensation energy we find that the strength of the Heisenberg exchange coupling is responsible for determining the magnitude of bose condensation energy, while the coupling between the spinon and holon pair channels induces the arch shaped structure of bose condensation energy as a function of hole doping concentration, as shown in Fig. 1 and Fig. 2(a).

We now seek a scaling relation between the condensation energy and the hole doping concentration $x$. 
The computed condensation energy for both the underdoped and overdoped regions is well fit by the following relation, $U = J |\Delta_f|^2 |\Delta_b|^2 \approx \alpha x^2 |\Delta_0|^2$ with $\alpha = 0.25/J$, as shown by the solid line in Fig.1.
Here the pseudogap is obtained to be $\Delta_0  = 2 J (1- x)^2 \Delta_f$ in our U (1) slave-boson theory\cite{LEE_SPEC}.
The pseudogap $\Delta_0$ below the superconducting temperature $T_c$ is observed to be the superconducting gap\cite{NORMAN,FENG,HARRIS,INO}. 
This is the gap also used by Momono et al. for their determination of the empirical relation for fitting the observed condensation energy which is given by $U = a N (0) x^2 \Delta_0^2$ with $a$, a constant. 
Our theory agrees well with the empirical relation obtained by Momono et al.\cite{MOMONO} in the sense that 
1. the condensation energy is well fit as a function of the hole density but not the electron density even in the overdoped region, and 2. there exists a strong doping ($x$) dependence of condensation energy.
However there exists a discrepancy in the power of $x$ (effectively $x^3$ dependence in their case).

For comparison of our predicted condensation energy with experiment, we find the maximum condensation energy at critical doping $x_c=0.1$ to be $U_{max} = 6.40 \times 10^{-5} t$ (per site) with the choice of $J/t = 0.2$.
With the use of $t=0.44 eV$\cite{HYBERTSEN} this value corresponds to $2.7 Joule/mole$.
This result is reasonably close to the observed value of condensation energy ($U_{max} \approx 3 Joule/mole$) of the single layer $La_{2-x}Sr_xCuO_4$ compound\cite{MOMONO}.
The maximum condensation energies are predicted to be $1.64 \times 10^{-4} t$ ($6.9 Joule/mole$) for $J/t=0.3$ and $3.13 \times 10^{-4}t$ ($13.2 Joule/mole$) for $J/t=0.4$, showing a rapid increase of $U$ with $J$.

To study correlations between the condensation energy and other physical quantities, in Fig. 2 we display the computed doping dependences of (a) condensation energy, (b) quasiparticle peak intensity and (c) the superfluid weight.
Fig.2(b) displays the doping dependence of SPR at ${\bf k}=(\pi,0)$, that is, the ratio of the spectral intensity of the quasiparticle peak to the total spectral intensity obtained from integration over an entire frequency range.
The SPR (normalized quasiparticle peak) is seen to increase with hole concentration in the underdoped region and reaches a maximum which occurs above the predicted optimal doping.
This is consistent with ARPES measurements\cite{FENG,HARRIS,INO}.
In Fig. 2(c), the doping dependence of the predicted superfluid weight is shown to exhibit a qualitatively similar behavior to the condensation energy, in that there exists an increasing trend of its magnitude in the underdoped region and a decreasing trend in the overdoped region.
In agreement with observations, the predicted pseudogap (spin gap) and, likewise, the spinon pairing order diminishes in the overdoped region.
Thus such decreasing trends of the condensation energy, spectral peak intensity and superfluid weight in the overdoped region are correlated with the diminishing trend of the spinon pairing order in the same region, thus resulting in the weak coupling to holon pair order as can be seen from the coupling term (the last term in Eq.(\ref{eq:holon_u1})).

In summary, in the present work we focused on a study of the doping dependence of the bose condensation energy and thus the relevance of bose condensation to both the spectral peak intensity at ${\bf k}=(\pi,0)$ and the superfluid weight.
It is shown that the bose condensation associated with the Cooper pairs results from the interplay between the charge and spin degrees of freedom, that is, the coupling between the spinon (spin) paring and holon (charge) pairing orders and that the condensation energy of $U$ at $T = 0 K$ is well fit by the relation, $U(0)  \approx \alpha x^2 |\Delta_0|^2$ with  $\Delta_0$, the spin gap at $0 K$; $x$, the hole doping concentration and $\alpha$ a constant.
In agreement with observations\cite{MOMONO} the condensation energy scales well with the hole density, $x$, but not with the electron density, $1-x$ for both the underdoped and overdoped regions.
Judging from the good correlations in the doping dependence between the condensation energy and spectral peak intensity at ${\bf k}=(\pi,0)$, the bose condensation is caused by the phase coherence of Cooper pairs formed largely from the anti-node region of d-wave symmetry which has the maximum gap energy.
This finding is qualitatively consistent with experimental observations\cite{FENG,INO}.
We find that there exists a universality of doping dependence in the bose condensation energy, the superfluid weight and the spectral intensity, by equally showing an arch shape feature for the entire range of hole doping concentration as shown in Fig. 1 and Fig. 2.
This indicates that these physical quantities are different manifestations of the same physical origin largely contributed from the momentum space near the antinodal points of the d-wave pairing order in high $T_c$ cuprates.

One(SHSS) of us acknowledges the generous supports of Korea Ministry of Education (HakJin Program 2002-2003) and the Institute of Basic Science Research (2002) at Pohang University of Science and Technology.
We are grateful to Professor M. Ido and Professor M. Oda for their valuable assistance for the present work.

\begin{minipage}[c]{9cm}
\begin{figure}
\vspace{0cm}
\epsfig{file=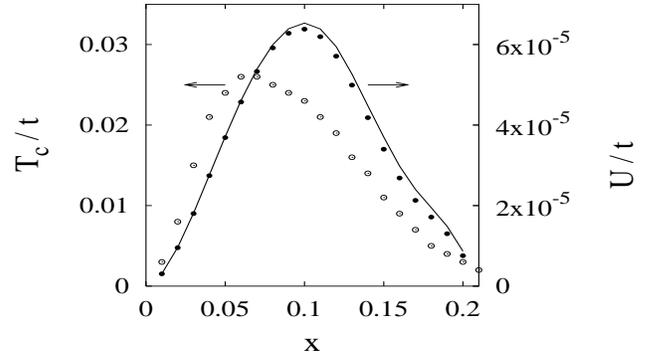,angle=0, height=5cm, width=9cm}
\label{fig:1}
\caption{
The doping dependence of condensation energy density at $T=0$ and of superconducting temperature $T_c$ for $J/t = 0.2$ with the U(1) slave-boson theory.  The open circles represent the predicted superconducting temperature and the solid circles, the predicted condensation energy.  The solid line represents a fit to the condensation energy density by the relation $U = J |\Delta_f|^2 |\Delta_b|^2 \approx \alpha x^2 |\Delta_0|^2$ with $\alpha=0.25/J$.
}
\end{figure}
 \end{minipage}

\begin{minipage}[c]{9cm}
\begin{figure}
\vspace{0cm}
\epsfig{file=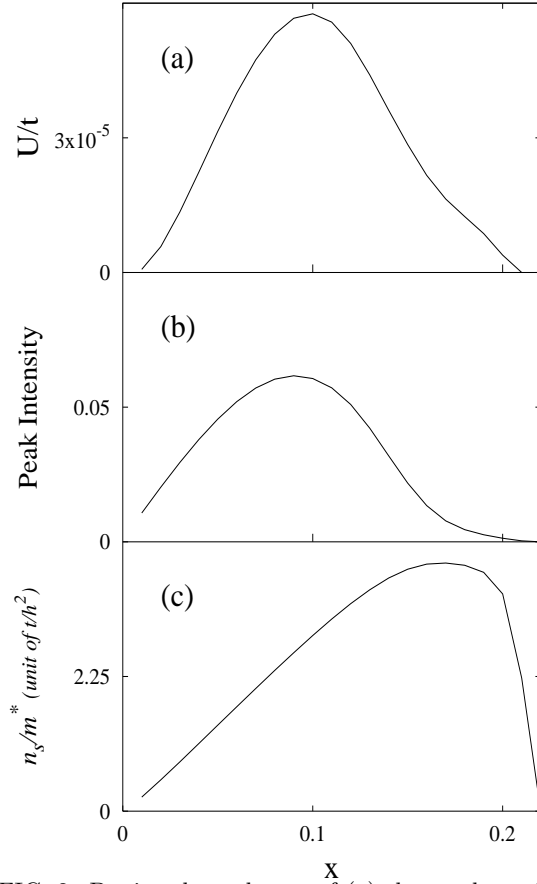,angle=0, height=12cm, width=8cm}
\label{fig:3}
\caption{
Doping dependences of (a) the condensation energy,  (b) the normalized peak intensity of spectral function and (c) the superfluid weight at $T/t=0.001$(equivalent to $5K$ with the use of $t=0.44eV$[18]) with $J/t=0.2$ for the U(1) slave-boson theory.
}
\end{figure}
 \end{minipage}


\end{multicols}
\end{document}